# Anisotropic Superconductivity of $Ca_{1-x}La_xFeAs_2$ ($x \sim 0.18$) Single Crystal

Wei Zhou, Jincheng Zhuang, Feifei Yuan, Xiong Li, Xiangzhuo Xing, Yue Sun, and Zhixiang Shi*

*Department of Physics and Key Laboratory of MEMS of the Ministry of Education, Southeast University, Nanjing 211189, China*

* Author to whom correspondence should be addressed. Email address: zxshi@seu.edu.cn

*Abstract*

Anisotropic superconducting properties including the upper critical field $H_{c2}$, thermal activation energy $U_0$, and critical current density $J_c$ are systematically studied in a large $Ca_{1-x}La_xFeAs_2$ single crystal ($x \sim 0.18$). The obtained $H_{c2}$ bears a moderate anisotropy $\gamma$ of approximately 2-4.2, located between those of "122" $Ba_{1-x}K_xFe_2As_2$ ($1 < \gamma < 2$) and "1111" $NdFeAsO_{1-x}F_x$ ($5 < \gamma < 9.2$). Both the magnitude of $U_0$ and its field dependence are very similar to those of $NdFeAsO_{1-x}F_x$, also indicating anisotropic superconductivity. Moreover, high and anisotropic $J_c$'s exceeding $10^5$ A/cm$^2$ have been calculated from the magnetization hysteresis loops, indicating the existence of strong bulk-dominated pinning in the present superconducting material.

The new finding of "112" type superconductor $Ca_{1-x}RE_xFeAs_2$ (RE = La, Pr) with onset transition temperature of approximately 43 K are attractive and significant[1,2]. Different from most iron-based superconductors (IBSs), the new superconductor crystallizes in a monoclinic structure with the space group $P2_1$ (No. 4)[1-4]. The alternative stacked layer between two adjacent FeAs layers consists of both an arsenic zigzag bond layer (As$^-$) and a rare earth element doped alkalis layer, which extends the layered structures for the iron arsenide family.

Superconductors with layered structures have attracted much attention owing to the usual occurrence of unconventional and high $T_c$ superconductivity. Extensive studies on cuprates indicate that a higher $T_c$ may be achieved by the more reduced dimensionality (or higher anisotropy) originating from the layered crystal structures with the quasi-two-dimensional (2D) electronic properties. Early work on IBSs also seemed to support the idea[5,6]. However, many subsequent research studies reported rather a low upper critical field $H_{c2}$ anisotropy ($\gamma$) in "122"- ($\gamma < 2$)[7] and "11" ($\gamma \sim$ 1.8)[8] -type IBSs, which seemed to contradict the above valid law for $T_c$ and anisotropy in cuprates. For the "1111" system with the highest $T_c$ in IBSs, a moderate anisotropy ($\gamma \sim 5$) was proved[9]. These $\gamma$ values reflect a relatively strong coupling strength between the charge reservoir layers and the conducting FeAs layers for IBSs, and further indicate that the underlying electronic structure seems remarkably three-dimensional. Besides $H_{c2}$, anisotropies of the thermal activation energy $U_0$ and the critical current density $J_c$ are also important parameters for exploring the relationship between superconductivity and dimensionality. Investigations of the anisotropic upper critical field $H_{c2}$, thermal activation energy $U_0$, and critical current density $J_c$ are highly desired for a newly found superconductor in the explorations of the superconductivity mechanism and application.

In this work, large and high-quality plate-like single crystals were synthesized. Both the highest zero-resistance temperature (also the sharpest transition width $\Delta T_c \sim$ 1.8 K) and the highest critical current density $J_c$ (over $10^5$ A/cm$^2$) were measured for our crystals. Benefitting from the high superconductivity performance, systematical anisotropy studies of the upper critical field $H_{c2}$, thermal activation energy $U_0$, and critical current density $J_c$ were performed for the first time. The anisotropic behaviors were discussed and compared in connection with the "122" and "1111" cases.

Single crystal growth and X-ray diffraction characterization are presented in the supplementary document (S1, S2). The main panel of Fig. 1 shows the temperature dependence of resistance (*R-T*). No antiferromagnetic-like (AFM) transition as in many un/under-doped IBSs can be observed[10]. The general shape of the *R-T* curve, similar to that of "1111"-type IBSs ($T_c \sim 55$ K)[11], does not show good linear behavior and also clearly deviates from standard Fermi-liquid theory[12]. The superconducting transition width $\Delta T_c$ is around 1.8 K and the high zero-resistance temperature $T_c^{zero}$ reaches 40.8 K. The $\Delta T_c$ is the smallest and the $T_c^{zero}$ is the highest among the existing reports, proving a high sample quality.[1,2,13,14]

Taking advantages of the good superconductivity performance, we have further investigated the superconductivity capability against magnetic field and its anisotropy. *R-T* curves with *H* applied both parallel to the *c*-axis (*H* || *c*) and *ab*-plane (*H* || *ab*) were measured (S3). The small shifts of $T_c$ for *H* up to 9 T in both directions indicate large $H_{c2}$ values, which is very advantageous for applications. In Fig. 2, the $H_{c2}$ phase diagram is established. Here the $H_{c2}$'s are extracted from the peaks in the $\partial R/\partial T$ curves. An upward curvature near $T_c$ in $H_{c2}(T)$ curves is observed, which seems to be a common feature in many IBSs and is usually explained on the basis of the two-band theory[15]. Nevertheless, the linear *T* dependences of $H_{c2}$ above 1 T satisfies well with the Werthamer-Helfand-Hohenberg (WHH) model[16]. According to the WHH model, the upper critical fields at zero temperature are roughly estimated to be $H_{c2}^c(0)$ = 39.4 T (*H* || *c*, $H_{c2}$ slope = 14 kOe/K) and $H_{c2}^{ab}(0)$ = 166.2 T (*H* || *ab*, $H_{c2}$ slope = 59 kOe/K). The corresponding coherence lengths are $\xi_c(0)$ = 6.9 Å and $\xi_{ab}(0)$ = 28.9 Å. The $H_{c2}$ anisotropy at zero temperature $\gamma(0)$ is 4.2. It should be pointed out that, the estimated $H_{c2}(0)$, $\xi(0)$, and $\gamma(0)$ should be modified for the possible existence of anisotropic paramagnetic effects as in many IBSs[7,17]. Therefore, we use the $\gamma$ value (~2.08) near $T_c$ for comparison with the other IBSs. On one hand, such a $\gamma$ value locates between those of "122"-type $Ba_{1-x}K_xFe_2As_2$ ($1 < \gamma < 2$)[7] and "1111"-type $NdFeAsO_{1-x}F_x$ ($5 < \gamma < 9.2$)[18, 19]. As mentioned above, $\gamma$ reflects directly the coupling strength between the charge reservoir layers and the conducting FeAs layers. That is, the distance $d$ between the adjacent FeAs layers plays an important role in determining $\gamma$ values. A long distance $d$ may lead to large $\gamma$ values. The $d$ values for $Ca_{1-x}La_xFeAs_2$ ($c$ = 10.35 Å) and $NdFeAsO_{1-x}F_x$ ($c$ = 8.56 Å)[18] are the same as the $c$-axis lattice constant $c$, while formula $d = c/2$ should be used for $Ba_{1-x}K_xFe_2As_2$ ($c$ = 13.30 Å)[4] owing to the existence of double FeAs planes in a single unit cell. Accordingly, $Ba_{1-x}K_xFe_2As_2$ exhibits the smallest anisotropy. Compared with $Ca_{1-x}La_xFeAs_2$, "1111"-type $NdFeAsO_{1-x}F_x$ bears a shorter FeAs distance $d$ but exhibits a larger $\gamma$. Possibly, the $\gamma$ values may also be affected by the other factors, such as the electrical conductivity of the reservoir layers ($(Nd(O, F))^{x+}$, $(Ba, K)^{x+}$, and $(Ca, La, As)^{x+}$. Here $x$ means the valence) stacked between FeAs planes. Because the reservoir layer $Nd(O, F)^{x+}$ in $NdFeAsO_{1-x}F_x$ bears poor electrical conductivity, aa slightly large anisotropy is reflected. On the other hand, compared with the extremely large $H_{c2}$ anisotropy in the "122"-type analogy superconductors $Ca_{1-x}RE_xFe_2As_2$ whose weak high-$T_c$ superconductivity is of filamentary nature[20-22], $\gamma$ in $Ca_{1-x}La_xFeAs_2$ is clearly small. In this view, the superconductivity origin in the new "112"-type compounds may be different from that of the filamentary superconductivity in $Ca_{1-x}RE_xFe_2As_2$. This speculation is also evidenced by the magnetic results discussed later. In the bottom-left inset of Fig. 2, temperature-dependent $\gamma$ near $T_c$ is shown. Similar to the trend in $NdFeAsO_{1-x}F_x$, $\gamma$ increases gradually from 2 to a value above 4 with decreasing temperature. This trend is opposite to that of $Ba_{1-x}K_xFe_2As_2$, which may be caused by multiband effects[7].

Generally, for the resistive transition in the presence of a magnetic field, the effect of thermally activated flow of vortices should be taken into account, which means that the temperature dependence of resistivity in this region is expected to be expressed by the Arrhenius plot[19], $\rho(T,H) = \rho_0 \exp[-U(T,H)/k_BT]$. Here, $U$ ($T$, $H$) is the activation energy of the flux flow (also known as the pinning potential/energy), $\rho_0$ is a constant, and $k_B$ is Boltzmann's constant. As can be seen in the upper-right and bottom-left insets of Fig. 3, the Arrhenius dependence holds about 2-3 decades. This linear behavior guarantees the suitable application of the formula $\rho(T,H) = \rho_0 \exp[-U_0(H)/k_B T]$.[19,23] The activation energy $U_0$ ($H$) [the main panel of Fig. 3] thus can be extracted from the slope of the curve $\log \rho(T,H)$ versus $1/T$. As can be seen, $U_0$ for both field orientations is approximately (3-4) × $10^3$ K at low fields. The field dependences of $U_0$ for $H \parallel ab$ and $H \parallel c$ are markedly different. For $H \parallel c$, $U_0$ shows a very weak dependence in the low-magnetic-field region ($H < 1$ T). A stronger power law decrease $U_0 \propto B^{-0.56}$ takes place for $H$ above 1 T. A similar feature of $U_0$ with double field-dependences was also found in cuprates and NdFeAsO$_{1-x}$F$_x$[19], which was explained by a transition from a single-vortex-dominated pinning to a small bundle pinning. In the state of single-vortex-dominated pinning, the overlap of vortices is negligible and the field dependence is weak. When $H$ is increased to a certain amount (noted as $H^*$) that results in significant overlaps of vortices, the pinning energy $U_0$ begins to be markedly suppressed. For $H \parallel ab$, a single power law $U_0 \propto B^{-0.29}$ holds through the field regions, which possibly results from a nearly negligible $H^*$. It is peculiarly interesting that, similar to NdFeAsO$_{1-x}$F$_x$, a slightly large $U_0$ is found for $H \parallel c$ in low field region in contrast with the result of Ba$_{1-x}$K$_x$Fe$_2$As$_2$.[24] We believe that the larger $U_0$ for $H \parallel c$ in Ca$_{1-x}$La$_x$FeAs$_2$ and NdFeAsO$_{1-x}$F$_x$ is caused by the anisotropic $H^*$, which is mainly determined by the anisotropic superconducting penetration depth $\lambda$ ($\lambda_{ab} < \lambda_c$). The large $\lambda_c$ leads to a nearly negligible $H^*$ for $H \parallel ab$. In the field region $H^* (H \parallel ab) < H < H^* (H \parallel c)$, the single-vortex pinning is dominated for $H \parallel c$, while the vortex pinning for $H \parallel ab$ has already stepped into the small bundle state. The weaker field dependence of $U_0$ in the single-vortex pinning state causes the larger value of $U_0$ ($H \parallel c$). However, in the high-field region, since $H_{c2}$ ($H \parallel ab$) is larger than $H_{c2}$ ($H \parallel c$), $U_0$ ($H \parallel c$) tends to decrease faster with the field than $U_0$ ($H \parallel ab$), resulting in a crossover of $U_0$. Overall, the anisotropic behavior of $U_0$ reflects a relatively two-dimensional superconductivity compared with that of $Ba_{1-x}K_xFe_2As_2$.

Magnetic hysteresis loops (MHLs) at 5 K with fields applied both parallel to $c$-axis ($H \parallel c$) and $ab$-plane ($H \parallel ab$) were also collected [Fig. 4 (a)]. The big and symmetric MHLs imply that the bulk pinning dominates in the crystal. Such large MHLs are observed for the first time in the present superconducting material. For a rectangular prism-shaped crystal of dimensions $2c < 2a < 2b$[25], there are three types of critical current $J_c^{x,y}$, where $x$ and $y$ refer to the directions of the current and magnetic field, respectively. When the magnetic field is applied along the $c$ side, the supercurrent density $J_c^{ab,c}$ is generated by the gradients of the Abrikosov vortices perpendicular to the $ab$ plane, which can be determined easily using the Bean model. Fig. 4(b) shows the calculated critical current density $J_c^{ab,c}$ under the Bean construction[25], $J = 20\Delta M / a(1-\frac{a}{3b})$ [Eq. (a)], where $\Delta M$ (unit: emu/cm$^3$) is $M_{down} - M_{up}$; $M_{down}$ and $M_{up}$ are magnetization when sweeping fields down and up, respectively. Here $a$ (unit: cm) and $b$ (unit: cm) are the redefined sample width and length ($a < b$) under a certain configuration of field orientation. As can be seen, the self field $J_c^{ab,c}$ at 5 K reaches a high value as $3.6 \times 10^5$ A/cm$^2$. Such a value is comparable to that of a well-annealed "11" crystal $Fe_{1+y}(Te, Se)$[8] whose critical current density is almost isotropic.

On the other hand, when the magnetic field is applied along the $a$ (or $b$) side, the situation becomes more complicated for the existence of two current densities with different directions, of which one is along the $ab$ plane ($J_c^{ab,ab}$) and another is across the planes ($J_c^{c,ab}$). Assuming that $J_c^{c,ab}$ is equal to $J_c^{ab,ab}$, the weighted average $J_c^{mean}$ for $H \parallel ab$ can be directly obtained using Eq. (a). The $J_c^{mean}$ values clearly exceeds $10^5$ A/cm$^2$. Even through $J_c^{mean}$ is frequently used in the $H \parallel ab$ case in many superconductors, the above assumption is somewhat unreasonable. By the method described in Ref. [26] and assuming $J_c^{ab,ab} = J_c^{ab,c}$, the relatively reliable $J_c^{c,ab}$ values are obtained, as shown in Fig. 4 (b). $J_c^{c,ab}$ reflects much smaller values than

$J_c^{mean}$, which is more reasonable for the reason that the coherence length $\xi_c$ (~ 6.9 Å) is smaller than the FeAs distance $d$ (~ 10.35 Å). Therefore, similar to the $H_{c2}$ anisotropy $\gamma$, a larger $J_c$ anisotropy is found for $Ca_{1-x}La_xFeAs_2$ than the nearly isotropic "122" compounds[26]. It should also be noted here that for the magnetic field applied both parallel to the c-axis and ab-plane, the values of $J_c$(5 K) slightly change against magnetic fields up to 6 T, revealing a magnetic-field-robust feature. The large and magnetic-field-robust $J_c$ could not come from interface or filamentary superconductivity. Actually, in the case of the filamentary superconductivity in $Ca_{1-x}RE_xFe_2As_2$, almost no superconducting MHLs can be measured[27].

Therefore, in terms of $T_c$ and superconductivity anisotropy, the "112"-type $Ca_{1-x}La_xFeAs_2$ ($T_c$ ~ 42.6 K) exhibits relatively moderate values, bridging the difference between "122" ($T_c$ ~ 38 K) and "1111" ($T_c$ ~ 55 K) iron-based compounds. Possibly, similar to cuprates, a high anisotropy may act as an important ingredient for the achievement of high transition temperatures.

In summary, large size $Ca_{0.77}La_{0.18}Fe_{0.90}As_2$ single crystals with the best superconducting performance are synthesized. The roughly estimated upper critical field $H_{c2}$ (0)'s using the WHH model in the present superconducting material reach values as high as 39.4 T and 166.2 T for the out-plane ($H \parallel c$) and in-plane ($H \parallel ab$) directions, respectively. The $H_{c2}$ anisotropy $\gamma$ near $T_c$ is 2.08. The anisotropic pinning potential $U_0$ shows comparable behaviors to that of "1111" $NdFeAsO_{1-x}F_x$. A high critical current density $J_c$ of over $10^5$ A/cm$^2$ is determined from magnetization hysteresis loops, indicating strong bulk-dominated pinning. The moderate anisotropy in the new "112"-type superconductor seems to bridge the difference between "122"- and "1111"-type IBSs.

**Acknowledgments**
This work was supported by the Natural Science Foundation of China, the Ministry of Science and Technology of China (973 project: No. 2011CBA00105) and the

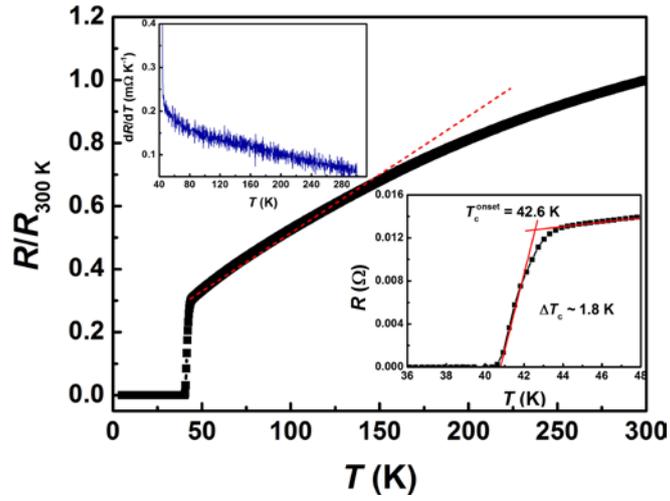

Fig. 1. Main panel: temperature dependence of normalized resistance ($R$-$T$) of $Ca_{0.77}La_{0.18}Fe_{0.90}As_2$. The red dashed line is a linear guide line. Bottom-right corner: enlarged view of superconducting transition. Upper-left corner: temperature dependence of $R$-$T$ derivative (d$R$/d$T$).

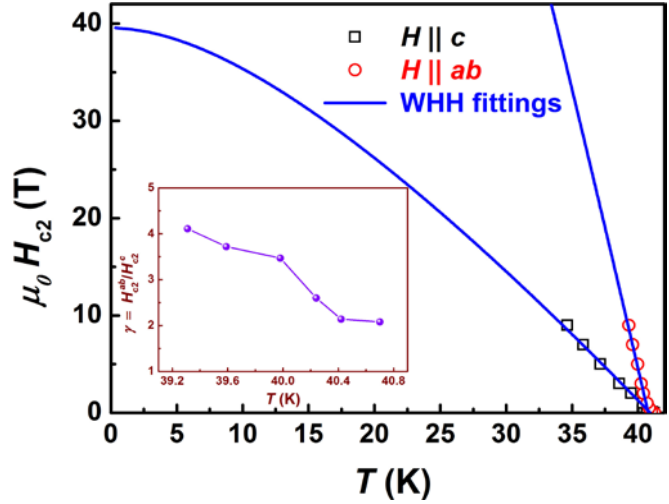

Fig. 2. $H_{c2}$ phase diagram obtained from $R$-$T$ curves under different magnetic fields using peak criterion. Solid lines are WHH fittings. Inset: temperature dependence of $H_{c2}$ anisotropy parameter $\gamma$ ($\gamma = H_{c2}^{ab}/H_{c2}^{c}$).

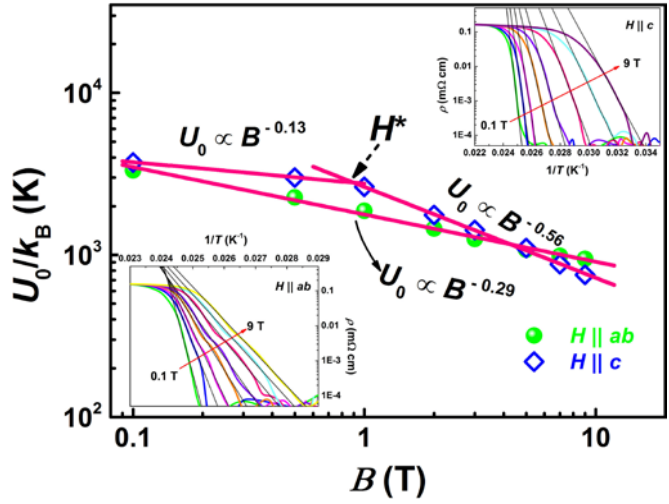

Fig. 3. Arrhenius plot for temperature-dependent resistivity at different magnetic fields parallel to $c$-axis ($H \parallel c$, upper-right corner) and $ab$-plane ($H \parallel ab$, bottom-left corner). Main panel: field dependence of activation energy $U_0$ for $H \parallel c$ and $H \parallel ab$. Solid lines are linear fittings in different field regions.

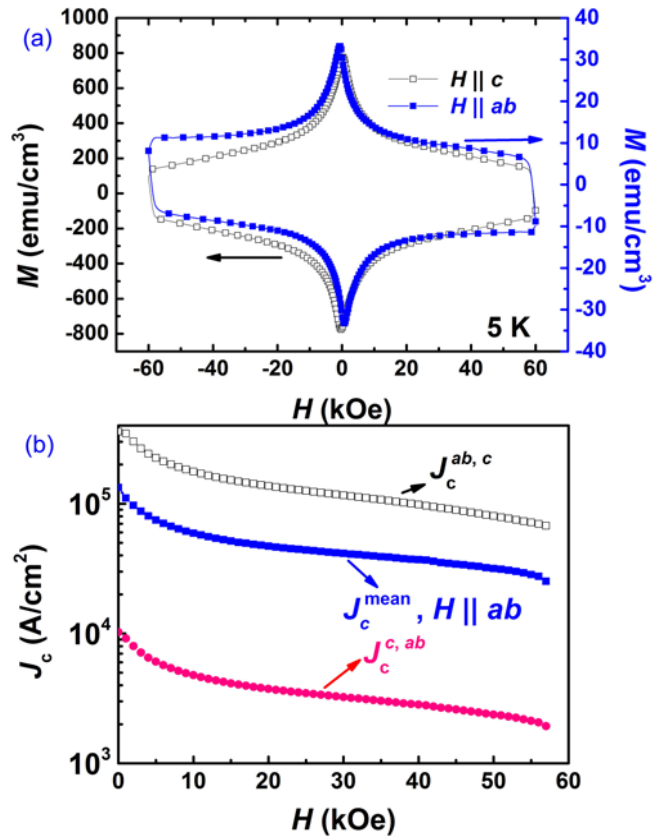

Fig. 4. Magnetic hysteresis loops (a) and critical current density $J_c$ (b) at 5 K with fields applied both parallel to $c$-axis ($H \parallel c$) and $ab$-plane ($H \parallel ab$). The $J_c$ calculation methods are stated in the main text.

# Anisotropic Superconductivity of $Ca_{1-x}La_xFeAs_2$ ($x \sim 0.18$) Single Crystal

Wei Zhou, Jincheng Zhuang, Feifei Yuan, Xiong Li, Xiangzhuo Xing, Yue Sun, and Zhixiang Shi*

*Department of Physics and Key Laboratory of MEMS of the Ministry of Education,*

*Southeast University, Nanjing 211189, China*

* Author to whom correspondence should be addressed. Email address: zxshi@seu.edu.cn

## Supplementary information

### S1. Single crystal growth

Single crystals with nominal composition $Ca_{0.9}La_{0.1}FeAs_2$ were grown using the similar flux method to the previous reports[1]. Small amount of CaO seems helpful for crystallization. We found that slightly large amount of the starting materials (~ 8 g) was very necessary for the crystal growth. Crystals obtained from our high-temperature box-furnace reaches size around 2 mm, which is the largest value in the existing reports[1-4]. The growth procedure for the relatively large crystals can be repeated easily. The actual La doping concentration $x$ used in this report was 0.18, which was determined by the average value of the energy dispersive X-ray spectroscopy (EDXS) measurement results of 50 random selected points. No O atom was detected in such method obtained crystal.

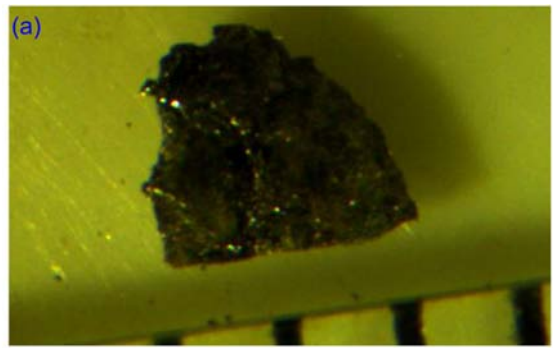

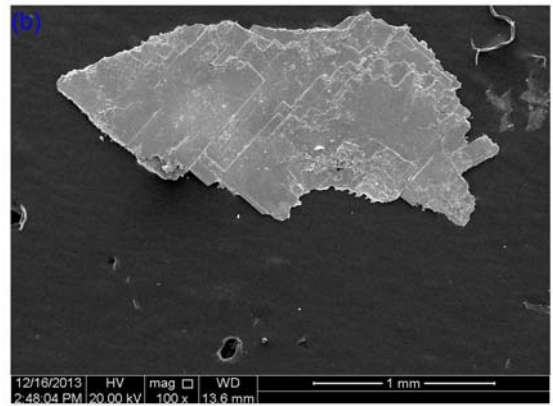

Fig. S1. (a) and (b) The topography of crystals employing optical microscope and scanning electron microscopy (SEM), respectively. The general size is around 2 mm.

## S2. X-ray diffraction (XRD) characterization

Single crystal X-ray diffraction (XRD) pattern, with Cu K$\alpha$ radiation from 5° to 90°, was performed. Only (00$l$) peaks which can be indexed with the space group $P2_1$ (No. 4) were observed, indicating good $c$-axis orientation. No trace of second phase even employing a semi-logarithm coordinate axis can be observed. The calculated $c$-axis lattice constant $c$ is 10.35 Å.

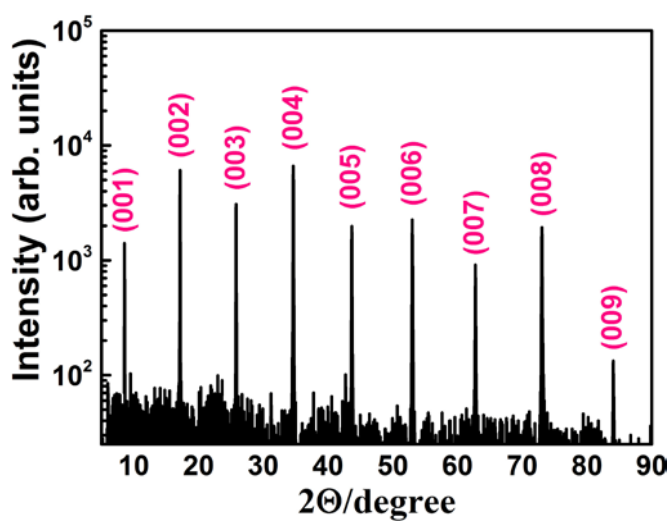

Fig. S2. Single crystal X-ray diffraction pattern plotted with a semi-logarithm coordinate axis for the single crystal. No trace of second phase was found.

## S3. Superconducting transition measurement under magnetic field

Temperature dependences of resistance under magnetic field $H$ applied both parallel to $c$-axis ($H \parallel c$) and $ab$-plane ($H \parallel ab$) were measured. Small shifts of $T_c$ less than 2 K were observed with field up to 9 T for $H \parallel ab$, indicating strong superconducting capability against magnetic field. The estimated $H_{c2}(0)$ values for $H \parallel c$ and $H \parallel ab$ using WHH model are 39.4 T and 166.2 T, repectivily.

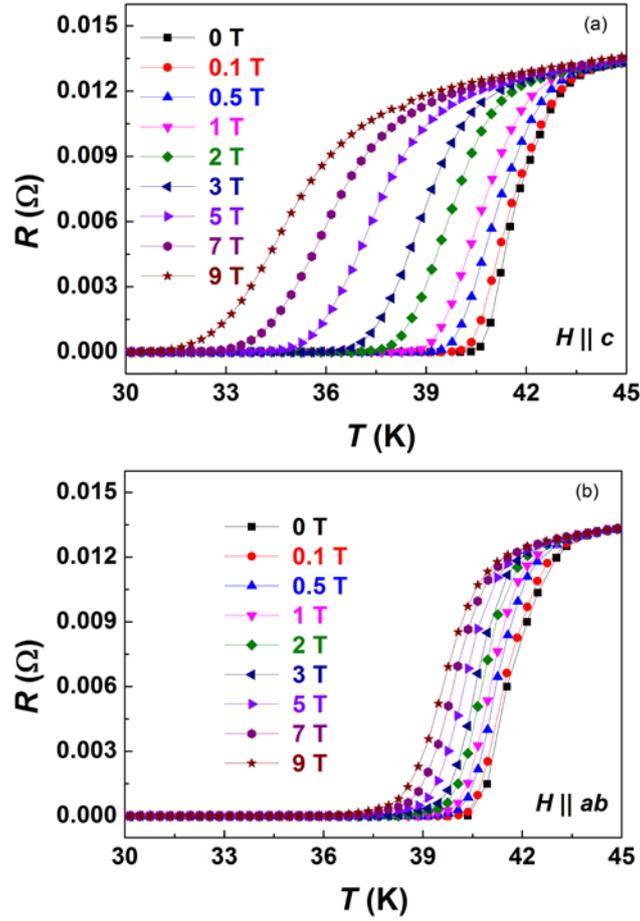

Fig. S3. Temperature dependences of resistance under magnetic fields applied parallel to *c*-axis [(a)] and *ab*-plane [(b)], respectively.